\documentclass[arXiv,twocolumn,
superscriptaddress]{revtex4}
\usepackage[intlimits]{amsmath}
\usepackage{amsfonts}
\usepackage{graphics}
\usepackage{bbm}
\usepackage{subfigure}
\usepackage[usenames]{color}
\usepackage{epstopdf,epsfig}
\usepackage{mathtools}
\usepackage{ulem}
\usepackage{soul}
\usepackage{bm}

\DeclarePairedDelimiterX\braket[2]{\langle}{\rangle}{#1 \delimsize\vert #2}

\usepackage{indentfirst}

\usepackage{graphicx}
\usepackage{amsmath}
\usepackage{amsfonts}
\newcommand\be            {\begin{equation}}
\newcommand\bea           {\begin{equation}\begin{array}l\displaystyle}
\newcommand\ee            {\end{equation}}
\newcommand\bes           {\begin{subequations}}
\newcommand\esu           {\end{subequations}}

\newcommand\eps           {\varepsilon}

\newcommand\p            {\partial}

\makeatletter
\newcommand{\vast}{\bBigg@{3.4}}
\newcommand{\Vast}{\bBigg@{5}}

\makeatother

\def\3pt#1#2#3{{\langle{#1}\vert{#2}\vert{#3}\rangle}}
\def\barray{\begin{eqnarray}}
\def\earray{\end{eqnarray}}

\usepackage{tikz}
\usetikzlibrary{calc}

\begin{document}

\title{Finite Temperature Off-Diagonal Long-Range Order for Interacting Bosons}

\author{A. Colcelli}
\affiliation{SISSA and INFN, Sezione di Trieste, Via Bonomea 265, I-34136 
Trieste, Italy}

\author{N. Defenu}
\affiliation{Institute for Theoretical  Physics, ETH  Z\"urich,
  Wolfgang-Pauli-Str. 27, 8093  Z\"urich, Switzerland}
\affiliation{Institute for Theoretical Physics, Heidelberg University, D-69120 Heidelberg, Germany}

\author{G. Mussardo}
\affiliation{SISSA and INFN, Sezione di Trieste, Via Bonomea 265, I-34136 
Trieste, Italy}

\author{A. Trombettoni}
\affiliation{Department of Physics, University of Trieste, Strada Costiera
  11, I-34151 Trieste, Italy}
\affiliation{CNR-IOM DEMOCRITOS Simulation Center, Via Bonomea 265, I-34136
  Trieste, Italy} 
\affiliation{SISSA and INFN, Sezione di Trieste, Via Bonomea 265, I-34136
Trieste, Italy}

\begin{abstract}
\noindent
Characterizing the scaling with the total particle number ($N$)
of the largest eigenvalue of the one--body density matrix ($\lambda_0$),
provides informations on the occurrence of the off-diagonal long-range
order (ODLRO) according to the Penrose-Onsager criterion.
Setting $\lambda_0\sim N^{\mathcal{C}_0}$, then
$\mathcal{C}_0=1$ corresponds to ODLRO.
The intermediate case,
$0<\mathcal{C}_0<1$, corresponds for translational invariant systems
to the power-law decaying of (non-connected) correlation functions and
it can be seen as identifying quasi-long-range order.
The goal of the present paper is to characterize
the ODLRO properties encoded in $\mathcal{C}_0$
[and in the corresponding quantities $\mathcal{C}_{k \neq 0}$
  for excited natural orbitals]
exhibited by homogeneous interacting bosonic systems at finite temperature
for different dimensions. We show that $\mathcal{C}_{k \neq 0}=0$ in
the thermodynamic limit. In $1D$ it is $\mathcal{C}_0=0$
for non-vanishing temperature, while in $3D$ $\mathcal{C}_0=1$
($\mathcal{C}_0=0$) for temperatures smaller (larger) than
the Bose-Einstein critical temperature. We then focus our attention
to $D=2$, studying the $XY$ and the Villain models, and the
weakly interacting Bose gas. The universal value of $\mathcal{C}_0$
near the Berezinskii--Kosterlitz--Thouless temperature $T_{BKT}$ is $7/8$.
The dependence of $\mathcal{C}_0$ on temperatures between $T=0$
(at which $\mathcal{C}_0=1$)
and $T_{BKT}$ is studied in the different models.
An estimate for the (non-perturbative) parameter $\xi$ entering the
equation of state of the $2D$ Bose gases, is obtained using low temperature
expansions and compared with the Monte Carlo result. We finally discuss
a double jump behaviour for $\mathcal{C}_0$, and
correspondingly of the anomalous
dimension $\eta$, right below $T_{BKT}$ in the limit of vanishing interactions.
\end{abstract}

\maketitle

\section{Introduction}

Off-diagonal long-range order in the one--body density matrix 
of Bose particles signals the appearance of Bose-Einstein condensation (BEC) in quantum systems.
This relation is established by the Penrose-Onsager criterion\,\cite{Penrose56}
which applies in all dimensions $D$ and at any temperature $T$, irrespectively of the presence 
of confining potentials. For its versatility, it constitutes a simple way
to determine whether a quantum Bose gas exhibits condensation and coherence
effects\,\cite{Anderson66, Huang95}. 

For $D\le2$ the Mermin-Wagner theorem\,\cite{Mermin66, Hohenberg67} ensures that --
for translational invariant systems with continuous symmetry such as interacting
bosons or $O(N)$ spin models with $N \ge 2$ --
no long-range order can be found at finite temperature.
Indeed, the theorem forbids the occurrence of spontaneous symmetry breaking
for $T>0$ in low dimensional systems,
where the symmetry of the Hamiltonian is always restored by the proliferation
of long--wavelength fluctuations, often called Goldstone modes. For a Bose gas
the Goldstone modes are represented by the phonons, which in $D=2$ destroy long-range order,
leaving low temperature superfluidity intact. In such a case, due to
the persistence of $U(1)$ symmetry, the equilibrium finite-temperature
average of the bosonic field operator
$\hat{\Psi}$ vanishes, due to the lack of phase coherence\,\cite{Huang95}. It is worth noting that a similar effect occurs in a wide range of systems, even if the Mermin-Wagner theorem does not strictly apply, when the scaling dimension of the bosonic order parameter $\hat{\Psi}$ becomes zero\,\cite{Defenu2016, Defenu2017Anis, Gori2017}. 

A compact way to define off-diagonal long-range order (ODLRO) is to introduce
the one--body density matrix ($1$BDM)\,\cite{Pitaevskii16}
\be
\label{rho_def}
\rho(\vec{x},\vec{y})\,=\,\left\langle \hat{\Psi}^\dag(\vec{x}) \hat{\Psi}(\vec{y}) \right\rangle\,,
\ee
where the field operator $\hat{\Psi}(\vec{x})$ destroys a particle at the point identified
by the $D$--dimensional vector $\vec{x}$. The $1$BDM, as an Hermitian matrix, satisfies
the eigenvalue equation
\be
\label{rho_eigeneq_def}
\int  \rho(\vec{x},\vec{y}) \, \phi_i(\vec{y}) \,d\vec{y}\,=\, \lambda_i \, \phi_i(\vec{x}) \,,
\ee
with the eigenvalues $\lambda_i$ being real. They denote the occupation number of
the $i$-th natural orbital eigenfunction $\phi_i$, with $\sum_i \lambda_i=N$, where $N$ is the total
number of particles. The occurrence of ODLRO (and therefore of BEC) is characterized by a
linear scaling of the largest eigenvalue $\lambda_0$ with respect to the total number
of particles $N$ in the system\,\cite{Penrose56,CNYang62}: $\lambda_0 \sim N$.

For a translational invariant system, the indices $i$ in Eq.\,\eqref{rho_eigeneq_def} are wavevectors, which are conventionally denoted by the vector $\vec{k}$. Introducing the scaling formula
\be
\label{lambda0_scaling_def}
\lambda_0 \sim N^{{\cal C}_0(T)} \,,
\ee
the Mermin-Wagner theorem implies that ${\cal C}_0(T)<1$ for $T>0$ and $D \le 2$,
so there is no ODLRO at finite temperature. One can show as well that ${\cal C}_0(T=0)=1$
for $D=2$ and ${\cal C}_0(T=0) < 1$
in $D=1$ (for the interacting case),
see Ref.\,\cite{Stringari95}. For a translational invariant system, absence of   
ODLRO, or equivalently of BEC, in $D=2$ at finite temperature
amounts to the following behaviour of the $1$BDM at large distances:
\be
\label{no-ODLRO2D}
\left\langle\hat{\Psi}^\dag (\vec{x})\hat{\Psi}(\vec{y})\right\rangle\,\xrightarrow{\text{$\left|\vec{x}-\vec{y}\right|\rightarrow\infty$}}\,\left\langle\hat{\Psi}(\vec{x})\right\rangle^* \left\langle\hat{\Psi}(\vec{y})\right\rangle\,=\,0\,.
\ee
The existence and regimes for BEC, \textit{i.e.} whether ${\cal C}_0=1$ or not, in various physical systems has been the subject of a remarkable amount of work. It would be therefore desirable to complete such analysis
with a systematic study of when ${\cal C}_0$ is smaller than $1$: In this case there is no ODLRO/BEC but nevertheless
the condition $0<{\cal C}_0<1$ implies that, in translational
invariant systems, the correlation
function $\langle \hat{\Psi}^\dag(\vec{x}) \hat{\Psi}(\vec{y}) \rangle$ have
a power-law decay. One may refer to this situation as
\textit{quasi-long-range order}.
A general classification of different behaviours of the correlation functions characterizing
different types of order is discussed in Ref.\,\cite{Yukalov91}.
Here, we find convenient to identify the ODLRO properties
in terms of the scaling with the particles number $N$ of the eigenvalues
$\lambda_k$ of the $1$BDM. 
Let's stress that, for a system of interacting bosons,
the index $\mathcal{C}_0$ may also
depend on the interaction strength and, moreover, one may expect that increasing the repulsion among the bosons, 
$\mathcal{C}_0$ gets dampened with respect to the weak interacting case, as seen 
explicitly in the $1D$ case at zero temperature\,\cite{EPL_nostro}.

In the present work, we are going to characterize ODLRO, and possible deviations
from it, in translational invariant bosonic systems interacting via short-range potential
in $1$-, $2$- and $3$-dimensions at finite temperatures.
With ``possible deviations'' we also mean a study of the behaviour of the index ${\cal C}_k(T)$, defined as
\be
\label{lambda0_scaling_def_k}
\lambda_k \sim N^{{\cal C}_k(T)} \,,
\ee
where $k \neq 0$. The study of $\mathcal{C}_{k\neq0}(T)$ gives an insight
about the possible \textit{quasi-fragmentation} of the system,
\textit{i.e.} how the particles occupy the other, $k\ne0$, states. 
Notice that in literature usually one refers to fragmentation when more
than one eigenvalue of the $1$BDM scales with $N$. So, one can refer
to the case in which at least two ${\cal C}_k$ are larger than zero
(and at least one is smaller than $1$) as a quasi-fragmentation.

The power-law behaviour in Eq.\,\eqref{lambda0_scaling_def} determines the leading scaling of the largest eigenvalue of the $1$BDM and, 
according to the Penrose-Onsager criterion, there is a BEC/ODLRO, \textit{i.e.} a macroscopic occupation of the lowest energy state, if $\mathcal{C}_0(T)=1$. 
There will be instead a mesoscopic condensate (i.e. quasi-long-range order), with a finite value for the condensate fraction $\frac{\lambda_0}{N}$ for finite values of $N$,
if $0<\mathcal{C}_0(T)<1$. In this case the condensate fraction of course vanishes for $N \to \infty$ but, even though the system is not a true BEC, one would observe
nevertheless a clear peak in the momentum distribution in an experiment with ultracold gases: The reason is that the number of particles which are typically used in these apparatus 
are of order $N\sim 10^3 - 10^5$, and therefore the condensate fraction $\frac{\lambda_0}{N}\sim\frac{N^{\mathcal{C}_0(T)}}{N}$
could be very close to the unity for $\mathcal{C}_0(T)$ close to $1$. For $\mathcal{C}_0(T)=0$
there will be no order at all, and the system behaves like a Fermi gas where
for all the eigenvalues we have $\lambda_i=1$, because of the Pauli principle.

The plan of the paper is the following. In Section \ref{II} we discuss the relation
between the $1$BDM and the momentum distribution, setting the notation for the Sections which follow.
The cases $D=3$ and $D=1$ are discussed respectively in Section \ref{3D} and \ref{1D}: these two cases 
provide the reference frame and the warming up for the discussion of the finite temperature
ODLRO properties of two--dimensional Bose gases in Section \ref{2D}. In Section \ref{2D} we also present a study of the ODLRO in the $XY$ and
the Villain Hamiltonians. Our conclusions are presented in Section \ref{concl}.

\section{Momentum distribution of homogeneous systems}
\label{II}

The advantage of studying how the largest eigenvalue scales with the number of particles,
instead of the large distance
behaviour of the $1$BDM, becomes evident once we define another important quantity: 
the momentum distribution. To introduce this quantity, let's initially consider the Fourier transform $\hat{\Psi}(\vec{k})$ of the 
field operator $\hat{\Psi}(\vec{x})$:
 $$\hat{\Psi}(\vec{k})\,=\,\frac{1}{(2\,\pi)^{D/2}}
\int \,d\vec{x} \,e^{i\,\vec{k}\cdot\vec{x}}\, \hat{\Psi}(\vec{x})$$
and the momentum distribution $n(k)$ given by 
\begin{eqnarray}
n(\vec{k})&\,=\,& \left\langle \hat{\Psi}^\dag(\vec{k}) \hat{\Psi}(\vec{k})\right \rangle\,,\label{mom_distr_def}\\
&\,=\,&\frac{1}{(2\,\pi)^D}\,\int\, d\vec{x} \int\, d\vec{y} \,e^{i\,\vec{k}\cdot(\vec{x}-\vec{y})}\,\left\langle \hat{\Psi}^\dag(\vec{x})\hat{\Psi}(\vec{y})\right\rangle\,.\nonumber
\end{eqnarray}
For a homogeneous system, $\rho(x,y)=\left\langle \hat{\Psi}^\dag(\vec{x})\hat{\Psi}(\vec{y})\right\rangle$ depends only on the distance among two points,
therefore writing the relative distance vector as $\vec{r}=\vec{x}-\vec{y}$,
we can rewrite $\rho(\vec{x},\vec{y})=\rho(\vec{r})$ and assume
$\rho(\vec{r})=\rho(\left|\vec{r}\right|)\equiv\rho(r)$. Passing to center of mass and relative coordinates,
since $\int d\vec{R} = L^D$ where $L$ denotes the size of the system
(\textit{e.g.} $L$ is the circumference of a ring in one--dimensional geometry), Eq. (\ref{mom_distr_def}) can be rewritten in an universal form
as 
\be
\label{mom_distr_generic}
n(\vec{k})\,=\,\left(\frac{L}{2\,\pi}\right)^D \int e^{i\,\vec{k}\cdot\vec{r}}\,\rho(r)\,d\vec{r}\,.
\ee
The integral in the right-hand-side depends of course on $D$. 
Notice that the momentum distribution peak is simply given by the integral of the $1$BDM 
\be
\label{mom_distr_peak}
n(\vec{k}=0)\,=\,\left(\frac{L}{2\,\pi}\right)^D \int \rho(r)\,d\vec{r}\,,
\ee
and, as expected, the large distance asymptotic of the density matrix determines the small momenta behaviour of the momentum distribution. 

For a homogeneous system the quantum number labeling the occupation of 
natural orbitals is clearly the wavevector $\big|\vec{k}\big|\equiv k$.
In particular, the Galilean invariance tells us that the effective single--particles
states $\phi_k(\vec{x})$ may be written as plane waves,
\textit{i.e.} $\phi_k(\vec{x})=\frac{1}{L^{D/2}}\,e^{i\,\vec{k}\cdot\vec{x}}$,
therefore from Eqs.\,\eqref{rho_eigeneq_def} and\,\eqref{mom_distr_generic} we obtain that the
dimensionless momentum distribution, $n(\vec{k})/L^D$, coincides with the eigenvalue
equation of the one--body density matrix, apart from a $(1/2\,\pi)^D$ factor. Therefore, for a homogeneous system
we have a one-to-one correspondence between the scaling of the eigenvalues of $\rho(r)$ and
the scaling of the dimensionless momentum distribution: 
\be
\label{relation_lambda_momdistr}
\lambda_k \,\sim\,N^{\mathcal{C}_k(T)}\,\sim\,\frac{n(k)}{L^D}\,.
\ee

The advantage of characterizing the different types of order in terms
of the exponent
$\mathcal{C}_k(T)$ instead of the large distance behaviour of the $1$BDM is now clear
and it stems from the fact that, in the experiments, it is easier to analyze the momentum distribution
peak instead of looking of what happens to $\rho(\vec{x},\vec{y})$ for very large
(ideally infinite) distances $\left|\vec{x}-\vec{y}\right|\rightarrow \infty$,
since one should discern with high precision if the $1$BDM is zero or not at large distances.

Since a complete closed form for the density matrix is not in general
available for all interaction strengths and temperatures, we cannot directly compute
the eigenvalues of $\rho(\vec{x},\vec{y})$ and then study their scaling with $N$.
In order to obtain this information we will use the following procedure. From the large distance
asymptotic behaviour of the $1$BDM, whose expression for different
configurations of the system is usually available in the literature, we first make it
a periodic function of period $L$ by adding terms which have the same scaling behaviour 
of the density matrix in the range $r\in\big[0,\frac{L}{2}\big)$,
and which represent the reflected parts in the range $\big(\frac{L}{2},L\big]$.
In this way we construct a fully symmetric and circulant matrix, whose eigenvalues
are known to be real, as required, since they represent the occupation numbers of the
system. Finally we perform the Fourier transform of this symmetrized
density matrix and obtain in this way the behaviour of the momentum distribution. Writing
$k$ as $k = \frac{2\,\pi}{L}\,l$ with $l\in \mathbb{N}$, the scaling of the largest eigenvalue of the $1$BDM can be identified just imposing $l=0$ and
 tracking its $N$ dependence. In this way, we are able to explicitly compute the exponent $\mathcal{C}_0(T)$ of the system. On the other hand, choosing $l \propto L$ the behaviour of the Fourier transform in the limit $N\rightarrow\infty$ at fixed density $n=N/L^D$ yields the expression for the exponents $\mathcal{C}_{k\neq0}(T)$ via Eq.\,\eqref{relation_lambda_momdistr}.

In the following, we aim to characterise the deviations form ODLRO at finite temperature
for homogeneous interacting Bose gases in different dimensions. After discussing the explicit expression for $\mathcal{C}_0$, we will also discuss the finite non-zero momenta landscape,
ruling out the possibility of having quasi-fragmentation in bosonic interacting systems
with repulsive interactions. Our findings provide a counterpart to the corresponding results
for fragmentation in macroscopically occupied states with eigenvalues scaling with $N$\,\cite{Nozieres95}.

\section{Three Dimensions}
\label{3D}
\begin{figure}[t]
\includegraphics[width=\columnwidth]{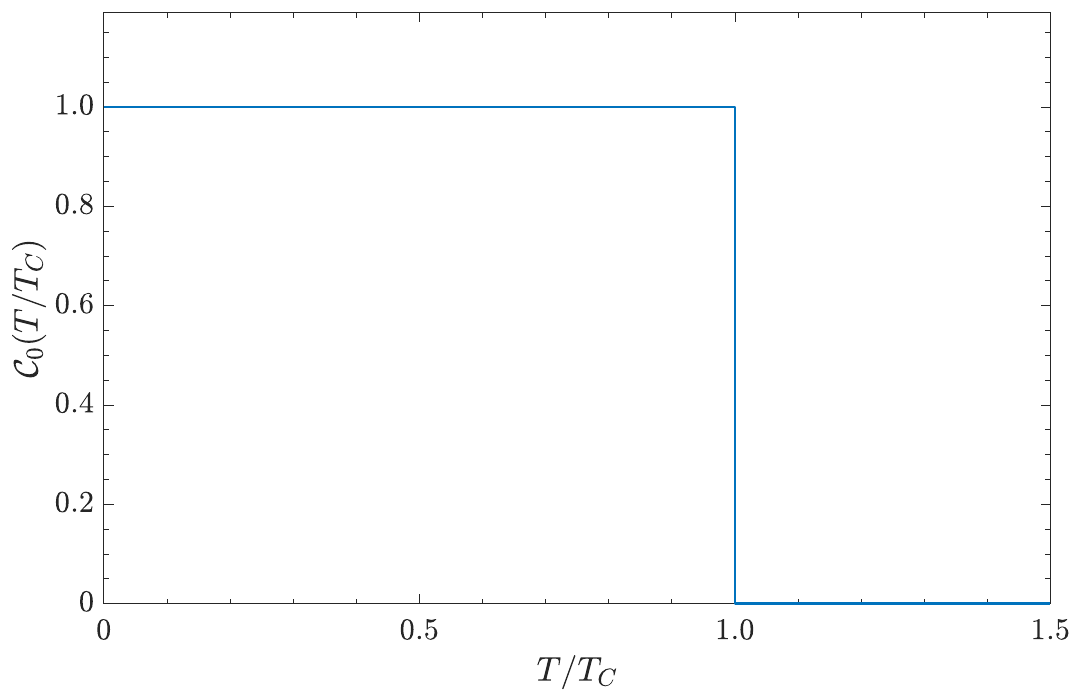}
\caption{Power $\mathcal{C}_0(T/T_C)$ with which the momentum distribution peak of a
  homogeneous three--dimensional Bose gas scales with respect to the total number of particles $N$ at different temperatures.}
\label{fig1}
\end{figure}

Let's begin with the case of a three--dimensional homogeneous Bose gas.
It is well known that, below the critical temperature $T_C$, a BEC takes place and the
lowest allowed state for the gas is then macroscopically occupied\,\cite{Pitaevskii16}.
This amounts to say that the momentum distribution of the system is
constituted by two parts: a non-singular part, relative to the occupation of the single particle states according to the Bose-Einstein distribution,
and a singular part $\propto N_0 \delta(\vec{k})$ which refers to the macroscopic occupation $N_0 \propto N$ of the lowest energy state, also called the condensate state. 
Therefore, at $T<T_C$ ODLRO are found and the exponent will be $\mathcal{C}_0=1$ in the condensed phase.
For temperatures above the critical $T_C$ there is no more condensation and
the singular part of the momentum distribution, \textit{i.e.} the Dirac delta peak,
disappears together with the system ordering. From all of these facts one can conclude that
\be
\label{C_3D_k=0}
\mathcal{C}_0(T)\,=\,
\begin{cases}
1, \,\,\,\,\,\, \text{for } T<T_C\\
0, \,\,\,\,\,\, \text{for } T>T_C
\end{cases}
\,,
\ee
as shown in Fig.\,\ref{fig1}.

In the weakly interacting Bose gas, one may use the Bogoliubov approximation\,\cite{Pitaevskii16} 
to obtain the scaling of the momentum distribution at $\vec{k}\neq 0$. Indeed, at this approximation level, the non-singular part of the momentum distribution at $T<T_C$ reads:
\be
\frac{n(\vec{k})}{L^D}\,=\,\frac{1}{(2\,\pi)^3}\frac{1}{e^{\eps(k)/k_B T}-1}\,,
\ee
where $\eps(k) = \sqrt{\frac{g\,n}{m} \,\hbar^2k^2+\left(\frac{\hbar^2 k^2}{2m}\right)^2}$ is
the Bogoliubov dispersion relation and $g=\frac{4\pi\hbar^2 a}{m}$ weights the interaction among
particles in terms of the s-wave scattering length $a$. Therefore for $\eps(k)/k_B T \gg 1$
we obtain:
\be
\frac{n(\vec{k})}{L^D}\,\simeq\,\frac{1}{(2\,\pi)^3}e^{-\frac{\hbar^2 k^2}{2 m k_B T}}\,\propto\, e^{-(l/L)^2}\,\propto\,N^0\,,
\ee
where in the second equality we used $k\propto l/L$, and in the last one we acknowledged that
$l\propto L$ in order to have a finite momentum $k$ in the thermodynamic limit. A similar procedure may be used to prove the absence of fragmentation also for $T>T_{C}$, yielding 
\be
\label{C_3D_kneq0}
\mathcal{C}_{k\neq 0}(T)\,=\,0\,,
\ee
at any temperature for the three--dimensional Bose gas. Notice that this result has been obtained
using Bogoliubov theory and it may not be applicable to gases with non-weak
interactions\,\cite{Mahan90,Capogrosso10}. However, since the exponents ${\cal C}$ are not expected to increase for larger interactions, one may reasonably conclude that this result is valid also for larger interactions.

\section{One Dimension}
\label{1D}

We now turn to the study of a one--dimensional homogeneous Bose gas\,\cite{GiamarchiBook,Cazalilla11}, within the framework of the
Lieb--Liniger model\,\cite{LiebLiniger63}, where the interaction between particles is represented by a repulsive $\delta$--potential.
The Lieb--Liniger Hamiltonian for $N$ bosons of mass $m$ then reads:
\be
\label{Ham_LL}
H=-\frac{\hbar^2}{2m}\sum_{i=1}^N\frac{\p^2}{\p x_i^2}+ 2 c \,
\sum_{i<j}\delta(x_i-x_j) \, ,
\ee
leading to the definition of the dimensionless coupling constant
\be
\gamma=\frac{2\,m\, c}{\hbar^2 \, n} \, , 
\ee
where $n = N/L$ is the density of the gas and $L$ is the size of the system
(with periodic boundary conditions this would be the circumference of the ring in which the
system is enclosed). As it is well known, the Lieb--Liniger model is exactly solvable by the Bethe
ansatz technique\,\cite{LiebLiniger63,Yang69} which provides an exact expression for the many--body eigenfunctions\,\cite{KorepinBook,GaudinBook}. Nevertheless a closed expression for the
$1$BDM for every coupling $\gamma$ and particle number $N$ is not known. One should then rely both 
on approximations\,\cite{Caux2006,Panfil2014} and numerical
approaches\,\cite{EPL_nostro,ABACUS}, which are suitable for working at large particle numbers.

At $T=0$, techniques coming from bosonization\,\cite{Haldane81,Cazalilla2004,Giamarchi2006} provides an
expression for the large distance behaviour of the density matrix for any values
of the interaction strengths\,\cite{KorepinBook,Calabrese2007}. In this case, the density matrix 
is written in terms of the dimensionless parameter called the Luttinger parameter,
which for Lieb--Liniger model reads $K=v_F/s$, where $v_F=\hbar \pi n/m$ is the
Fermi velocity and $s$ is the sound velocity of the Lieb--Liniger gas,
which depends on $\gamma$ and can be obtained via
Bethe ansatz\,\cite{Citro2011, Minguzzi2017}. At leading order, the large distance
asymptotic of the $1$BDM reads:
\be
\label{1BDM_1D_T0}
\frac{\rho(r)}{n}\,\simeq\,\frac{B_0}{\left(n\,r\right)^{1/2K}}\,,
\ee
where $B_0$ is a numerical prefactor\,\cite{PanfilCaux2012}. Symmetrizing
its expression in order to retrieve periodic boundary conditions, and then performing
the integral between $0$ and $L$, we get access to the dimensionless momentum distribution
peak scaling
\barray
\nonumber
\frac{n(k=0)}{L}\,&=&\,\frac{n^{1-1/2K}\,B_0}{2\,\pi}\left[\int_0^{L/2} \frac{dr}{r^{1/2K}}+\int_{L/2}^L \frac{dr}{(L-r)^{1/2K}}\right]\\
\label{mom_distr_peak_symm_powerlaw}
& &\propto\,N^{1-1/2K}\,,
\earray
which implies:
\be
\label{C_1D_T0}
\mathcal{C}_0(T=0, \gamma)\,=\,1-\frac{1}{2K(\gamma)}\,,
\ee
in agreement with Ref.\,\cite{EPL_nostro}. We verified that Eq.\,\eqref{mom_distr_peak_symm_powerlaw} also
holds also if we symmetrize the density matrix according to the formula
\be
\label{symmetrizing_with_sin}
\rho(r)\,\simeq\,\frac{n^{1-1/2K}\,B_0}{\left[\frac{L}{\pi}\sin\left(\frac{\pi\,x}{L}\right)\right]^{1/2K}}\,.
\ee
Notice that $\mathcal{C}_0(T=0,\gamma)$ depends only on $\gamma$ through the Luttinger parameter,
\textit{i.e.} it depends on the ratio $c/n$ and not on the interaction strength and the
density separately. The power $\mathcal{C}_0(T=0,\gamma)$ varies between $1$ for
$\gamma\rightarrow0$, to the value $1/2$ obtained for the Tonks--Girardeau
gas\,\cite{Lenard1964,Forrester2003,PRA_nostro}. For very small values of the
interaction parameter, say $\gamma\approx10^{-4}$, one gets $\mathcal{C}_0\approx0.99$,
which is very close to unity. Therefore the condensate fraction, $\lambda_0/N$,
for finite number of particles can be large and this could be seen in experiments
with $Rb$ atoms (when this occurs, one can say it is
in presence of a mesoscopic condensate).

Since $\lambda_0$ scales less than linearly with $N$ and at the same time we should have $\sum_k \lambda_k=N$,
in principle we could expect that at least for small values of $k$ there may exist some ${\cal C}_{k \neq 0}$ different from zero. 
However, as we are going to show in the following, this is not the case 
in the thermodynamic limit. To obtain the behaviour of the momentum distribution at non-zero
momenta, we have to perform the Fourier transform, for which we get:
\barray
\nonumber
\frac{n(k)}{L}\,&\propto&\, \int_0^{L/2} \frac{e^{ikr}}{r^{1/2K}}\,dr+\int_{L/2}^L \frac{e^{ikr}}{(L-r)^{1/2K}}\,dr\\
\nonumber 
& &\propto\,L^{1-1/2K}\,_1F_2\left(\frac{1}{2}-\frac{1}{4K};\frac{1}{2},\frac{3}{2}-\frac{1}{4K};-\frac{\pi^2 l^2}{4}\right)\,,
\earray
where $_1F_2\left(a;b_1,b_2;c\right)$ is the generalized hypergeometric function and
we used the fact that $k\,L=2\,\pi\,l$ with $l\in\mathbb{N}$. Expanding the hypergeometric
function for large $l$ and keeping only the leading term, we obtain
\be
\frac{n(k)}{L}\,\propto\,L^{1-1/2K} \, l^{-1+1/2K} \,\propto\,N^0\,,
\ee 
where in the last equality we used the fact that $l$ needs to grow like $L$
in the thermodynamic limit in order to have a fixed finite momentum $k$. Therefore the power
$\mathcal{C}_{k\neq0}$ for the one--dimensional gas at zero temperature and any interaction
strength is simply vanishing
\be
\label{C_1D_T0_kneq0}
\mathcal{C}_{k\neq0}(T=0,\gamma)\,=\,0\,,
\ee
and there is no fragmentation of the mesoscopic condensate.
The same result can be found also using Eq.\,\eqref{symmetrizing_with_sin}.

In Ref.\,\cite{EPL_nostro} it was verified that the largest eigenvalue of
the density matrix indeed scales with the exponent in Eq.\,\eqref{C_1D_T0}
by directly computing $\rho(r)$ using an interpolation method, which allows to get a
simple expression for the density matrix valid at any distance and interaction strengths.
The power-law scaling shows very good agreement, confirming that the method
sketched above to get access to the power $\mathcal{C}$ is correct.
We have then used the same interpolation scheme to get access to the $N$ dependence of
the $k\neq0$ eigenvalues of the $1$BDM\,\cite{PhD}. Apart from oscillations at small
particle numbers arising from a competition between the growth of $l$ and $L$,
for very large values of $N$ the eigenvalues $\lambda_{k\neq0}$ saturates
and the power $\mathcal{C}_{k\neq0}$ is indeed vanishing, confirming our theoretical prediction. 

In the finite temperature ($T\neq0$) case, several results are available for the
asymptotic behaviour of the density matrix of the Lieb--Liniger
gas\,\cite{Its1989,Its1992,PatuKlumper2013}. In Ref.\,\cite{PatuKlumper2013}
an expression for the $1$BDM as a sum of exponential functions is given in the form:
\be
\label{1BDM_1D_T}
\frac{\rho(r)}{n}\,=\,\sum_{i} \bar{B}_i \,e^{-\frac{r}{\xi\left[\bar{v}_i\right]}}\,,
\ee 
with $\bar{B}_i$ are distance independent amplitudes and $\xi\left[\bar{v}_i\right]$
the correlation length (shown to be always positive), depending on the temperature-dependent
functions $\bar{v}_i$ defined in Ref.\,\cite{PatuKlumper2013}, where it is also shown that
the result in Eq.\,\eqref{1BDM_1D_T} reduces to Eq.\,\eqref{1BDM_1D_T0}
in the $T=0$ case, as it should. We may now take the Fourier transform
of the symmetrized version of Eq.\,\eqref{1BDM_1D_T}, and obtain 
\barray
\nonumber
\frac{n(k)}{L}\,&=&\,\frac{1}{2\,\pi}\sum_i \bar{B}_i \left(\int_0^{L/2} \,e^{-\frac{r}{\xi\left[\bar{v}_i\right]}}\,dr+\int_{L/2}^L \,e^{-\frac{L-r}{\xi\left[\bar{v}_i\right]}}\,dr\right)\\
\nonumber 
& & \propto\,e^{-\frac{L}{\xi\left[\bar{v}_i\right]}}\,,
\earray
where the last proportionality is valid both at zero and non-zero momentum $k$.
Since $L = N/n$, analyzing the $N$ leading dependence only, we have that for $N\rightarrow \infty$
the dimensionless momentum distribution is just a constant for any $k$, leading to the finite temperature result:
\be
\label{C_1D_T}
\mathcal{C}_k(T\neq0,\gamma)\,=\,0\,,
\ee
which indicates complete absence of ordering.

\begin{figure}[t]
\includegraphics[width=\columnwidth]{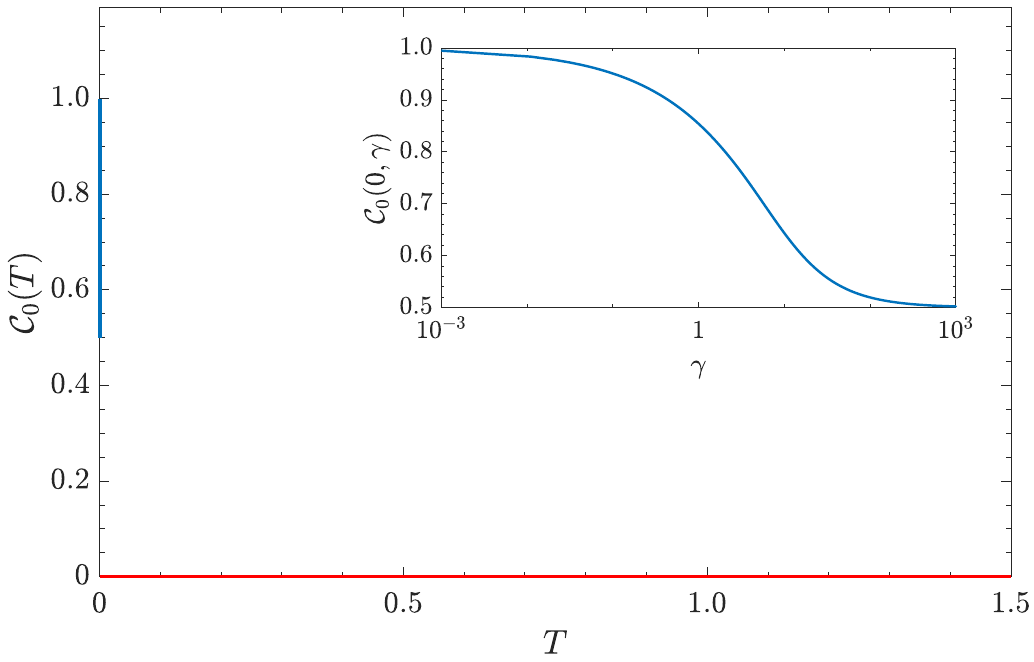}
\caption{Exponent $\mathcal{C}_0(T)$ with which the largest eigenvalue of
  the $1$BDM of a homogeneous one--dimensional Bose gas scales with respect
  to the total number of particles $N$ at different temperatures. Only for $T=0$ one has
  that $\mathcal{C}_0$
  is non-vanishing and depends on the dimensionless interaction parameter $\gamma$ via
  (\ref{C_1D_T0}), as shown in the inset.}
\label{fig2}
\end{figure}

In Fig.\,\ref{fig2} we summarize the behaviour of the exponent $\mathcal{C}_0$
for a homogeneous one--dimensional Bose gas for different temperatures.
An inset shows the relation between $\mathcal{C}_0$ and the interaction parameter $\gamma$
in the zero temperature case, \textit{i.e.} Eq. (\ref{C_1D_T0}).

\section{Two Dimensions}
\label{2D}

Properties of two--dimensional systems stand on their own and are 
between those of $1D$ -- where ${\cal C}_0$ vanishes at finite temperature --
and of $3D$ models -- where ${\cal C}_0=1$ below the BEC critical temperature. As discussed in the introduction, no ordinary phase transition takes place in $2D$, due to the lack of ODLRO. However, $2D$ systems often feature
the BKT topological phase transition named after Berezinskii, Kosterlitz and Thouless
who first discussed it in the two--dimensional XY
model\,\cite{Berezinskii70,Thouless73,Kosterlitz74}. This transition is related to the presence
of vortex and anti-vortex spin configurations at finite temperatures.
At low $T$, below the BKT temperature $T_{BKT}$, vortex and anti-vortex pairs with vanishing
total winding numbers (neutrality condition)
are present in the system and the correlation function between two distant spins
decay as power-law, indicating a phase with quasi-long-range order, also called BKT phase.

A simple estimate of $T_{BKT}$ in the $XY$ model is the Peierls value
$T_{BKT}=\frac{\pi J}{2 k_B}$\,\cite{LebellacBook}, where $J$ is
the interaction strength among the spins. In the low-temperature BKT regime
the only relevant configurations are the spin waves and the spin-wave approximation shall describe the system properly. As the temperature increases, the
presence of free vortices with non-vanishing winding numbers becomes energetically favoured,
and, therefore, vortices and anti-vortices may unbind from each other.
For temperatures above $T_{BKT}$, the presence of such topological excitations
destroy the quasi-long-range order and the correlation functions
become exponentially decaying\,\cite{LebellacBook,SimanekBook,OrtizBook}.
An important statistical model used to approximatively describe the
two--dimensional XY model is the one proposed by Villain \cite{Villain75,LebellacBook}.
While in the XY model the spin waves interact with the vortices, in the
Villain model the spin waves are decoupled from the vortices degrees of freedom,
making its Hamiltonian simply quadratic. Both models have the same topological
characteristics and they belong to the same universality class, as one can see 
from the critical behaviour of the anomalous dimension $\eta$ of the two systems.
The Villain model well describes the low temperature phase of the XY model,
since the Hamiltonian is essentially constituted by two decoupled harmonic oscillators terms,
one for the spin waves and one for the vortices. Notice that the Villain model can be used
both as a model \textit{per se} and also as a convenient way to approximate the XY 
model\,\cite{KleinertBook}.

Let pause here to comment on the qualitative similarity of the low dimensional ($D=1$ and $D=2$)
systems studied in this work.
In the thermodynamic limit at low temperatures,
both for the one-- and two--dimensional cases,
the systems can be described by field theoretical models with Hamiltonians made up
of two decoupled harmonic oscillators terms. These quadratic Hamiltonians are the
Luttinger liquid and the Villain Hamiltonian for the one-- and two--dimensional cases respectively.
Therefore bosonization in $D=1$ systems plays to a certain extent a similar role as
the spin wave approximation in $D=2$ systems, both of them describing systems with quasi-long-range order in the low temperature phase and absence of order above their
critical temperatures (which is vanishing in $D=1$). Nevertheless, the
phase transitions that characterize the models are for short-range models
intrinsically different in the one--
and two--dimensional cases. In $D=2$ this phase transition is related to the
formation of single independent topological excitations, which cannot happen in $D=1$ geometries.
Moreover in one dimension there is no phase transition at all at finite $T$,
since the quasi-long-range order is limited to the zero temperature limit.

Let us analyze the BKT phase transition in terms of the exponent
${\cal C}_0$.
At the BKT critical point the two--points correlation function scales as\,\cite{MussardoBook}
\be
\label{1BDM_2D_Bose}
\rho(r)\,\sim\,\frac{1}{r^{D-2+\eta}}\,,
\ee
where $\eta$ is the anomalous dimension critical exponent, that
depends on the system under consideration. What is universal is the value
at $T=0$, for which $\eta(T=0)=0$, and that at $T=T_{BKT}$, which is
given by: $\eta(T=T_{BKT})=1/4$\,\cite{Nelson77}.
The behaviour of $\eta$ between $0$ and $T_{BKT}$ is not universal.

From the knowledge of the behaviour of the anomalous dimension -- that will be discussed below --
one can find an expression for the power $\mathcal{C}_0$ with
which the dimensionless momentum distribution peak scales. One has
\barray
\nonumber
\frac{n(k=0)}{L^2}\,&=&\,\frac{1}{2\,\pi} \lim_{L \to \infty} \left[\int_0^{L/2} \frac{dr}{r^{\eta-1}}+\int_{L/2}^L \frac{dr}{(L-r)^{\eta-1}}\right]\\
\label{mom_distr_peak_2D}
& & \propto\,L^{2-\eta}\,,
\earray
where we symmetrized the density matrix in Eq.\,\eqref{1BDM_2D_Bose} in the radial coordinate variable $r$,
passing to polar coordinates and performing the trivial integration over the azimuth angle.
Since fixing the density $n=\frac{N}{L^2}$ in the large particle number limit implies
that $L\propto\sqrt{N}$, then we can extract the power $\mathcal{C}_0(T/T_{BKT})$ with which
the largest eigenvalue of the $1$BDM scales, and it reads:
\be
\label{C_2D_general_eta}
\mathcal{C}_0\,=\,1-\frac{\eta}{2}\,.
\ee 
Notice that for the XY and Villain models the condensate fraction $\frac{\lambda_0}{N}$ is
the magnetization density of the spin system and therefore Penrose-Onsager ODLRO manifests in
a complete magnetization of the system, while having $\mathcal{C}_0=0$
is equivalent to say that there exist no correlation and order between the spin variables.

Since the value of the anomalous dimension for such systems at the critical temperature
is equal to $1/4$, one has 
\begin{equation}
  \mathcal{C}_0\left(T=T_{BKT}\right) = \frac{7}{8}\,,
  \label{7_8}
  \end{equation}
  and ${\cal C}_0$ jumps to zero for $T>T_{BKT}$,
  reflecting the universal jump for the superfluid stiffness\,\cite{Nelson77}.
  A study of small corrections (found to be $\approx 0.02\%$)
  to the Nelson-Kosterlitz jump of the superfluid stiffness
  is in Refs.\,\cite{Prokofev00,Hasenbusch05}.  
  Using spin wave approximation, one finds that at $T=0$ there is ODLRO and therefore
  $\mathcal{C}_0(0)=1$. Notice that at $T=0$ ODLRO is allowed because there
  is no entropy contribution to the free energy of the system and the Mermin-Wagner theorem does not apply. 

\subsection{Villain model}
  
In the case of the square lattice planar Villain model, one expects that the anomalous
dimension should be of the form $\eta_V \simeq \frac{k_B T}{2\pi A}$ at low temperatures,
since the theory is quadratic and the spin wave approximation shall apply everywhere, in particular very close to the critical point, where vortex configurations become relevant. The value for
$A$ will be provided in the following.
Villain\,\cite{Villain75} proposed a correction term to account for vortex contributions to the anomalous dimension close to the critical point. Assuming that the interaction between the vortices can be neglected, this correction yields\,\cite{Villain75}:
\be
\label{eta_Villain1}
\eta_V\,=\,\frac{k_B T}{2\pi A}+ \pi^2 k_B T\,\frac{e^{-\pi^2 A /k_B T}}{\pi A-2 k_B T}\,.
\ee
According to the renormalization group, the value for the critical temperature of the
Villain models is found to be\,\cite{JankeNather93}
\be
\label{crit_temp_Villain}
\frac{k_B T_{BKT}}{A}\,=\,\frac{1}{0.74}\simeq 1.351\,,
\ee
which coincides with the result obtained from the high precision Monte Carlo simulation
performed in Ref.\,\cite{JankeNather93} up to $L=512$ lattice sites. Substituting
Eq.\,\eqref{crit_temp_Villain} into Eq.\,\eqref{eta_Villain1}, we have an estimate for
the behaviour of the anomalous dimension of the square lattice Villain model in terms of the
dimensionless ratio $T/T_{BKT}$, which reads:
\be
\label{eta_Villain2}
\eta_V(T/T_{BKT}) \,=\,{\cal A}\,\frac{T}{T_{BKT}}+\frac{\pi^2}{2}\frac{e^{-{\cal B}\,\frac{T_{BKT}}{T}}}
    {\left(-1+{\cal D}\,\frac{T_{BKT}}{T}\right)}\,,
\ee
where ${\cal A}\approx 0.215$, ${\cal B} \approx 7.304$ and ${\cal D}\approx 1.162$.

\begin{figure}[t]
\includegraphics[width=\columnwidth]{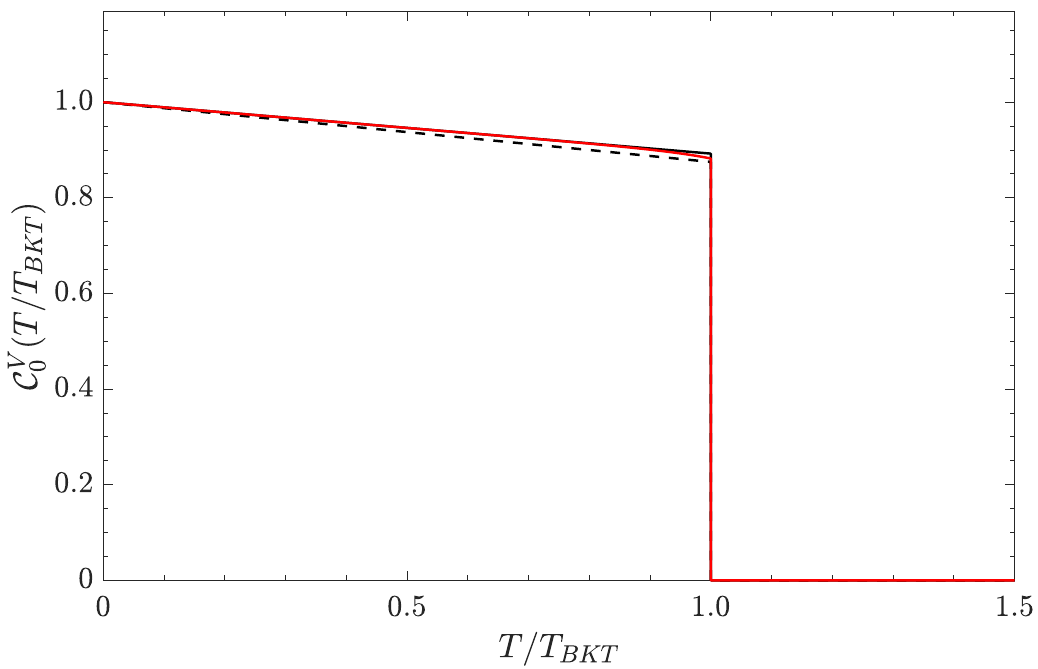}
\caption{$\mathcal{C}_0^{V}(T/T_{BKT})$ vs $T/T_{BKT}$ for the Villain square lattice model.
  The red intermediate solid line represents the predicted value for $\mathcal{C}_0$ using
  Eq. (\ref{eta_Villain2}) in Eq. (\ref{C_2D_general_eta}), while the black solid and
  dashed lines represent respectively the low temperature behaviours gives respectively
  by Eqs. (\ref{C_Villain_lowT1}) and (\ref{C_Villain_lowT2}).}
\label{VillainFigure}
\end{figure}

Introducing Eq.\,\eqref{eta_Villain2} into Eq.\,\eqref{C_2D_general_eta}, one obtains
the results plotted as the red intermediate solid line in Fig.\,\ref{VillainFigure}.
Notice that according to the approximation in Eq.\,\eqref{eta_Villain1},
 one has $\eta_V(T=T_{BKT}) \simeq 0.236$, \textit{i.e.} $\mathcal{C}_0^V (1)= 0.882$, with ``$V$''
referring to the Villain model. This
result differs from the one coming from Monte Carlo simulations \cite{JankeNather93},
$\eta_V = 0.2495 \pm 0.0006$, for about $5\%$. Low temperature predictions
for the exponent $\mathcal{C}_0^V (T)$ may be formulated in two ways:
\begin{enumerate}
\item Disregarding the second term in the
right-hand-side of Eq. (\ref{eta_Villain1}), which may be safely neglected in the low temperature regime
at $T \ll T_{BKT}$\,\cite{Villain75}, which yields, via Eq.\,\eqref{C_2D_general_eta},  
\be
\label{C_Villain_lowT1}
\mathcal{C}_0^V(T/T_{BKT})\,\simeq \, 1-\frac{1}{2}\left( \frac{T}{2\, \pi\, T_{BKT}} \cdot \frac{1}{0.74}\right)\,
\ee
with $T_{BKT}$ obtained by Monte Carlo simulations [see Eq.\,\eqref{crit_temp_Villain}]. 
\item Using the Peierls argument $\frac{k_B T_{BKT}}{A} = \frac{\pi}{2}$, one has: 
\be
\label{C_Villain_lowT2}
\mathcal{C}_0^V(T/T_{BKT})\,\simeq \, 1-\frac{1}{2}\left( \frac{T}{T_{BKT}} \cdot \frac{1}{4}\right)\,.
\ee
\end{enumerate}
These two behaviours are reported as black solid and dashed lines, respectively, in Fig.\,\ref{VillainFigure}.
Notice from the plot that the low-$T$ behaviour of Eq. (\ref{C_Villain_lowT1}) is good
even in region close to $T_{BKT}$, where the corrective term introduced by Villain starts
to play a role. The predictions of (\ref{C_Villain_lowT2}), which at variance do not
take into account the effect of vorticies, do not match with the same accuracy
with the expected results already from $T \approx 0.5 T_{BKT}$.

\subsection{XY model}

For the two--dimensional classical XY model, the critical temperature has been evaluated
using Monte Carlo techniques obtaining\,\cite{GuptaPRL,GuptaPRB,Schultka,Komura12}:
\be
\label{crit_temp_XY}
\frac{k_B T_{BKT}}{J}=0.893 \pm 0.001\,,
\ee
while recent approximate, semi-analytical functional renormalization group (FRG)
results give $k_B T_{BKT}=\left(0.94 \pm 0.02\right) J$\,\cite{Defenu2017}.
The anomalous dimension is
found to be equal to $$\eta_{XY}=\frac{k_B T}{2\pi J_s(T)}\,,$$
where $J_s(T)$ is the superfluid (or spin) stiffness of the model, and
has been recently calculated for the XY model in a square lattice in Ref.\,\cite{Maccari2017} using
simulations up to $256$ lattice sites. 

Therefore we may now compute the $k=0$ Fourier transform of the spin--spin correlation
function as in Eq.\,\eqref{mom_distr_peak_2D}. 
Similarly to Eq.\,\eqref{C_2D_general_eta}, one has 
\be
\label{C_2D_XY}
\mathcal{C}_0^{XY}(T)\,=\,1-\frac{\eta_{XY}}{2}\,.
\ee 

Using the Villain approximation we can obtain an expression for the behaviour of
the anomalous dimension for the XY model. The Villain approximation, indeed, is based on the fact
that there exist a (non-exact) map between the interaction parameter $A$ and the
spin--spin interaction parameter $J$, which relates the Villain Hamiltonian
to the XY model\,\cite{Villain75}. This mapping reads
\be
\label{relation_AconJ}
\frac{A}{k_B T}\,=\,-\frac{1}{2} \left\{\ln\left[\frac{I_1\left(\frac{J}{k_B T}\right)}{I_0\left(\frac{J}{k_B T}\right)}\right]\right\}^{-1}\,,
\ee
where $I_n(x)$ are the modified Bessel functions of the first kind of degree $n$.
We may therefore substitute this expression into the approximation given in Eq.\,\eqref{eta_Villain1}. We find:
\barray
\nonumber
\eta_{XY}&=&-\frac{1}{\pi} \ln\left[\frac{I_1\left(\frac{J}{k_B T}\right)}{I_0\left(\frac{J}{k_B T}\right)}\right]+\frac{\pi^2}{2} e^{\frac{\pi^2}{2} \left\{\ln\left[\frac{I_1\left(\frac{J}{k_B T}\right)}{I_0\left(\frac{J}{k_B T}\right)}\right]\right\}^{-1}} 
\\ \nonumber 
& & \left\{ -1 + \frac{\pi}{\pi +4 \ln\left[I_1\left(\frac{J}{k_B T}\right) \right]- 4\ln \left[I_0\left(\frac{J}{k_B T}\right)\right]}\right\}\,.
\\ \label{eta_XY_VillainApprox}
\earray
Using the mapping of Eq.\,\eqref{relation_AconJ}, the Monte Carlo results of Ref.\,\cite{JankeNather93}
for the critical temperature of the Villain model, \textit{i.e.} Eq.\,\eqref{crit_temp_Villain},
translates into $$\frac{k_B T_{BKT}}{J}=0.842\,,$$
which is pretty close to the Monte Carlo results
of Refs.\,\cite{GuptaPRL,GuptaPRB,Schultka,Komura12} reported in Eq. (\ref{crit_temp_XY}). 
The equation which relates $A$ to $J$ seems then to be reliable within a $\approx 6\%$ accuracy
even very close to the critical point. 

Similarly to what we have done for the Villain model,
a low temperature prediction can be made by neglecting the second term in
the right-hand-side of Eq.\,\eqref{eta_XY_VillainApprox}. Using Eq.\,\eqref{C_2D_XY} we get:
\be
\label{C_XY_lowT1}
\mathcal{C}_0^{XY}(T)\,\simeq\,1+\frac{1}{2\pi} \ln\left[\frac{I_1\left(\frac{J}{k_B T}\right)}{I_0\left(\frac{J}{k_B T}\right)}\right]\,.
\ee
On the other hand, one can also employ the low-temperature expansion results: $\frac{J_s(T)}{J}\simeq1-\frac{k_B T}{4J}$, which is known to be consistent with several approaches, such as
self-consistent harmonic approximation\,\cite{Pires96}, Monte Carlo simulations\,\cite{Hasenbusch}
and FRG\,\cite{Defenu2017}. This procedure leads to the expression:
\be
\label{C_XY_lowT2}
C_0^{XY}(T/T_{BKT})\,\simeq\,1-\frac{1}{\pi}\frac{T/T_{BKT}}{\frac{4J}{k_B T_{BKT}}-T/T_{BKT}}\,.
\ee

\begin{figure}[t]
\includegraphics[width=\columnwidth]{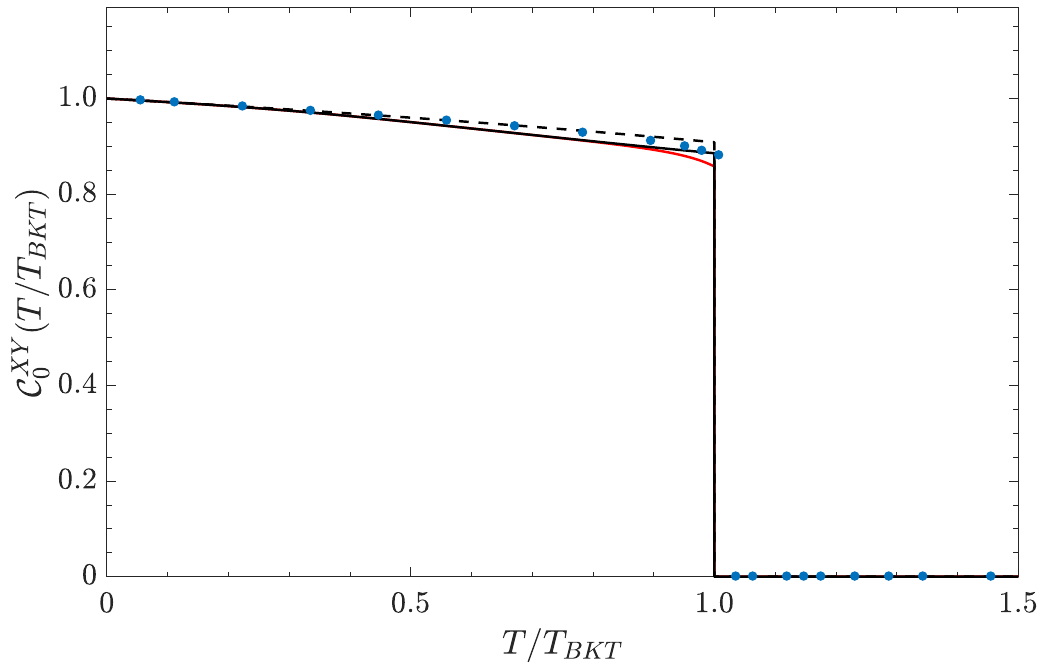}
\caption{$\mathcal{C}_0^{XY}(T/T_{BKT})$ vs $T/T_{BKT}$. Blue points are the numerical values
  of $\mathcal{C}$ obtained from Eq. (\ref{C_2D_XY}) with anomalous dimension
  $\eta_{XY}=\frac{T}{2\pi J_s(T)}$ and using the superfluid stiffness
  results
  of Ref.\,\cite{Maccari2017}. The universal jump from $\mathcal{C}_0(T_{BKT}) = \frac{7}{8}$
  to $\mathcal{C}_0(T>T_{BKT}) = 0$ is evident. The bottom red solid line comes from
  the Villain prediction Eq. (\ref{eta_XY_VillainApprox}).
  Finally the black solid and dashed lines represent the low temperature
  predictions of Eqs.\,\eqref{C_XY_lowT1} and\,\eqref{C_XY_lowT2} respectively.}
\label{fig3}
\end{figure}

In Fig.\,\ref{fig3} we report as blue points the behaviour of (\ref{C_2D_XY})
for $\eta_{XY}=\frac{k_B T}{2\pi J_s(T)}$ with respect to the dimensionless quantity $T/T_{BKT}$
obtained using the results of Ref.\,\cite{Maccari2017}.
The bottom red solid line represents the Villain prediction given in Eq.\,\eqref{eta_XY_VillainApprox}
with $T_{BKT}$ given by Eq.\,\eqref{crit_temp_XY},
while the black solid and dashed lines represent the low temperature behaviours
in Eqs.\,\eqref{C_XY_lowT1} and\,\eqref{C_XY_lowT2}, respectively. Fig.\,\ref{fig3} confirms the validity of the low temperature expansion in Eq.\,\eqref{C_XY_lowT2} in the range $T\in [0, 0.8 T_{BKT}]$, while
the Villain prediction in Eq.\,\eqref{C_XY_lowT2} remains reliable up to $T_{BKT}$.

\subsection{Bose gas}

Under certain conditions a two--dimensional Bose gas can be mapped
onto the XY model and from this mapping one can derive
the decay of correlation functions and the 
 ordering type of the bosonic system\,\cite{Prokofev2001,Svistunov2002,Trombettoni2005,Hadzibabic2011}. Indeed, when density fluctuations are strongly suppressed the effective low--energy
Hamiltonian of a two--dimensional Bose gas is equivalent to the
continuous version of the Hamiltonian of the XY model on the lattice.
The BKT phase of the XY model corresponds then to the superfluid state of
the Bose gas and quasi-long-range order is present. Above the critical
temperature the normal state appears and superfluidity breaks down. This abrupt change of phase is characterized by a
universal jump of the superfluid density (stiffness), which switches between its low temperature value $\rho_s=\frac{2m^2 k_B T}{\pi \hbar^2}$ to $\rho_s=0$ for $T>T_{BKT}$\,\cite{Nelson77,Prokofev00}. 

In Refs.\,\cite{Kane67,Popov72} it has been shown that the asymptotic behaviour of
the $1$BDM of a two--dimensional weakly interacting Bose
gas at finite temperatures scales as 
\be
\label{asympt_1bdm_bose}
\rho(r)\,\sim\,\frac{1}{r^{\frac{m^2\,k_B T}{2\pi\hbar^2 \rho_s}}}\,,
\ee
where $\rho_s$ is the superfluid density of the gas. The superfluid density
of the system assumes the form\,\cite{Svistunov2002}:
\be
\label{rs}
\rho_s\,=\,\frac{2m^2 k_B T}{\hbar^2\pi}\,f(X)\,,
\ee
where $X = \frac{\hbar^2 (\mu-\mu_c)}{m\,k_B T U}$ measures the
distance from the critical point, with $\mu$ the chemical potential and the
critical value $\mu_c$ given by:
\begin{equation}
  \mu_c = \frac{m\,k_B T U}{\hbar^2 \pi}\,
  \ln\left(\frac{\hbar^2 \xi_\mu}{m\,U}\right)\,.
  \label{mu_c}
\end{equation}  
The function $f(X)$ in Eq.\,\eqref{rs} is a dimensionless universal function, which has been  numerically determined
in Ref.\,\cite{Svistunov2002}. The variable $U$ appearing in $X$ is the interparticle
interaction strength, so that $\frac{m U}{\hbar^2}\ll 1$ and
$X \gg 1$ correspond to the
weakly interacting limit. While, the constant $\xi_\mu$ appearing in Eq.\,\eqref{mu_c} is given by
$\xi_\mu = 13.2 \pm 0.2$\,\cite{Svistunov2002}.

Applying the same procedure used for the Villain and the XY models,
we obtain the following exponent $\mathcal{C}_0$ for the scaling
of the dimensionless momentum distribution peak with respect to the number of
particle of the two--dimensional Bose gas:
\be
\label{C_2D_BoseGas}
\mathcal{C}_0^{\rm Bose}(X)\,=\,1-\frac{1}{8 f(X)}\,.
\ee
The jump of the superfluid stiffness $\rho_s$ at criticality implies that
$f(X)$ will jump from $0$ to $1$ at $X=0$, \textit{i.e.} at
the critical point. Therefore, the exponent $\mathcal{C}_0$
will jump from the universal value $\frac{7}{8}$ to $0$ at
the critical BKT temperature. The relation between the exponent $\mathcal{C}_0$ and the
ratio $T/T_{BKT}$ is constructed from the expression\,\cite{Svistunov2002}:

\be
\label{T/TBKT_2D_bose}
\frac{T}{T_{BKT}}(X)\,=\,\frac{1}{1+2\,\pi \lambda(X)/\ln(\hbar^2 \xi/mU)}\,,
\ee
where $\lambda(X)=\left[X+\theta(X)-\theta_0 \right]/2$
with $\theta(X)$ found via numerical simulations for system sizes up to $512$
in Ref.\,\cite{Svistunov2002}. The (non-perturbative) constant $\xi$ in Eq.\,\eqref{T/TBKT_2D_bose} is given by\,\cite{Svistunov2002}: 
\begin{equation}\xi=380\pm3\,,
  \label{xi}
  \end{equation}
and $\theta_0=\frac{1}{\pi}\ln\left(\frac{\xi}{\xi_\mu}\right)$
is then found to be
$\theta=1.07\pm 0.01$.

\begin{figure}[t]
\includegraphics[width=\columnwidth]{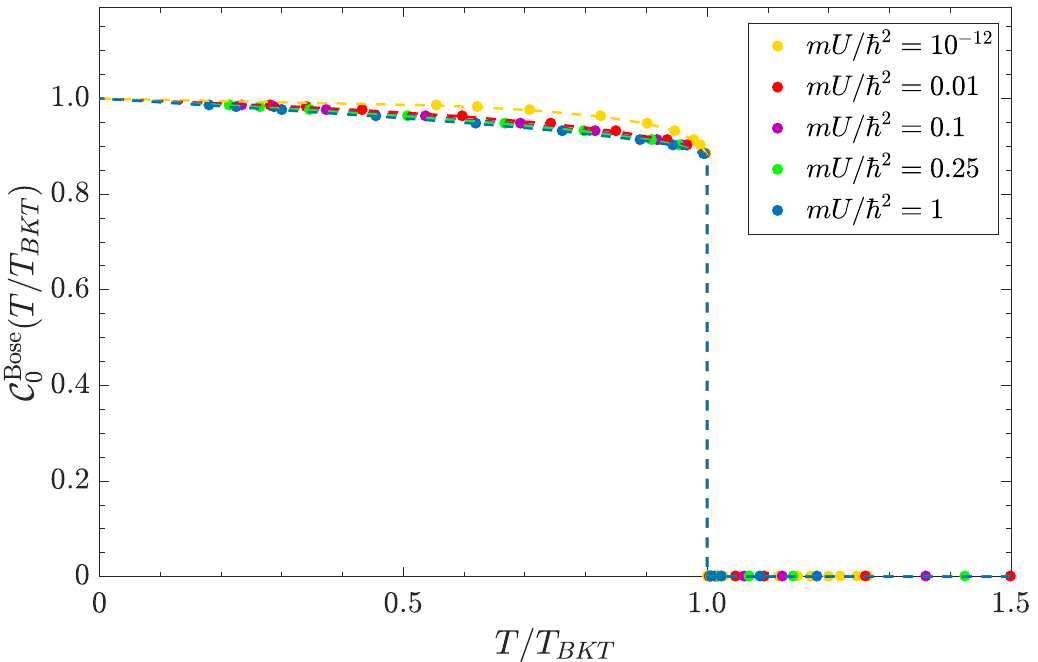}
\caption{$\mathcal{C}_0^{\rm Bose}(T/T_{BKT})$ vs $T/T_{BKT}$ for different
  interactions $\frac{mU}{\hbar^2}$. Points are the numerical value
  of $\mathcal{C}_0$ obtained from numerical simulations performed
  in Ref.\,\cite{Svistunov2002}, while dashed lines are drawn as guide for the eyes.
  In each case the universal jump from $\mathcal{C}_0(T_{BKT}) = \frac{7}{8}$ to
  $\mathcal{C}_0(T>T_{BKT}) = 0$ is evident.}
\label{fig4}
\end{figure}

Knowing the relation between $T/T_{BKT}$ and $X$ and the relation between $\mathcal{C}_0^{\rm Bose}$ and $X$,
we can then track down the dependence of the exponent ${\cal C}_0$ with
which the dimensionless momentum distribution peak scales with the
number of particles $N$ for different temperatures. We report its
behaviour in Fig.\,\ref{fig4} for different values of the interaction $U$.

An important comment about Fig.\,\ref{fig4} is that
in the limit of the dimensionless interaction parameter
$\frac{mU}{\hbar^2} \rightarrow 0$, the exponent $\mathcal{C}_0$ tends to be
closer (with respect to higher values of $U$)
to the
unity up to temperatures closer to $T_{BKT}$. In other words, the smaller is
$U$, the closer to $1$ is ${\cal C}_0$ at fixed $T/T_{BKT}<1$.
Going further close to $T_{BKT}$ from below, 
the decrease to the value $\frac{7}{8}$ happens abruptly
for $\frac{mU}{\hbar^2} \simeq 0$ at $T\simeq T_{BKT}$. Since
$\mathcal{C}_0$ has to be $7/8$ at $T=T_{BKT}$, this is associate to a kind of
{\it double jump} occurring for $T \to T_{BKT}^{-}$ for $U \to 0$,
since in this limit
$\mathcal{C}_0$ reaches a value different from (and larger than)
$7/8$ coming from low temperature/large-$X$ expansion that we are going to
shortly introduce,
then it abruptly jumps from this value
to $7/8$ and then jumps from $7/8$ to $0$. More comments on the double jump
occurrence are below.

Finally, it is worth noting that the values for
$\frac{mU}{\hbar^2}=1$, reported in Fig.\,\ref{fig4}, are out of the validity range for
the weak interacting gas. Then, the mean field arguments of Ref.\,\cite{Svistunov2002}
cannot be applied anymore, and one should take into account
quantum fluctuations.  

Low temperature predictions may also be formulated, similarly to what
we did for the Villain and XY models, but with some subtleties
to be worked out. In the low $T$ regime
(\textit{i.e.} far from the critical point),
it is $X\rightarrow \infty$ and the function $\theta(X)$ satisfies\,\cite{Svistunov2002}:
\be
\label{eq_theta(X)}
\theta(X)-\frac{1}{\pi}\ln\theta(X)\,=\,X+\frac{1}{\pi}\ln(2\xi_\mu)\,,
\ee
which is a transcendental equation admitting two values for
$\theta$ for a single value of $X$. These two solutions can be
distinguished in terms of the behaviour of $\theta(X)$ for
$X\rightarrow \infty$. The first set is the one having a 
vanishing vaue of $\theta(X\rightarrow\infty)$ and it is given by:
\be
\theta(X\rightarrow \infty)\,=\,\frac{e^{-\pi X}}{2\xi_\mu}\,,
\ee
which is the solution of
$-\frac{1}{\pi}\ln\theta(X)=X+\frac{1}{\pi}\ln(2\xi_\mu)$ and as well as
a solution of Eq. (\ref{eq_theta(X)}) for $X\rightarrow\infty$. This first set
is not interesting for us and we look for a function
$\theta(X)$ which diverges for large $X$. This represents the second set
of solutions and one has
\be
\label{solution_theta0}
\theta^{(0)}(X\rightarrow \infty)\,=\,X+\frac{1}{\pi}\ln(2\xi_\mu)\,,
\ee
which is the zero-th order solution of Eq.\,\eqref{eq_theta(X)}
without the logarithmic term in the left hand side. In the low $T$ regime
one may also write\,\cite{Svistunov2002}
\be
\label{f_lowT0}
f^{(0)}(X\rightarrow\infty)\,=\,\frac{\pi}{2}\theta^{(0)}(X)-\frac{1}{4}\,=\,\frac{2\pi X+2\ln(2\xi_\mu)-1}{4}\,,
\ee
where the last identity follows from Eq.\,\eqref{solution_theta0}.
Reminding that $\lambda(X)=\left[X+\theta(X)-\theta_0 \right]/2$ and
using Eq.\,\eqref{solution_theta0}, one has an expression also for the
$\lambda(X)$ function in the low temperature regime at the zero-th order
of approximation:
\be
\label{lambda_lowT0}
\lambda^{(0)}(X\rightarrow \infty)\,=\,X+\frac{1}{2\pi}\ln\left[\frac{2(\xi_\mu)^2}{\xi}\right]\,.
\ee
Therefore, substituting into Eq.\,\eqref{T/TBKT_2D_bose}, one can write
an expression for $X$ (at the zero-th order in terms of the variable
$T/T_{BKT}$) reading:
\be
\label{X_lowT0}
X^{(0)}\,=\,-\frac{1}{2\pi}\ln\left[\frac{2(\xi_\mu)^2}{\xi}\right]+\frac{1}{2\pi}\ln\left(\frac{\hbar^2\xi}{mU}\right)\left(\frac{T_{BKT}}{T}-1\right)\,.
\ee
Finally, inserting Eq.\,\eqref{X_lowT0} in Eq.\,\eqref{f_lowT0}, we may
substitute the equation for $f^{(0)}(X\rightarrow\infty)$ into
Eq.\,\eqref{C_2D_BoseGas} to obtain an analytical expression for the exponent
$\mathcal{C}_0^{\rm Bose}$ at low temperatures: 
\barray
\nonumber
\mathcal{C}_0^{\rm Bose\,(0)}(T/T_{BKT})\,&\simeq&\,1+\frac{1}{2}\Bigg[1-\ln(2\xi)\\
\nonumber
& &-\ln\left(\frac{\hbar^2 \xi}{mU}\right)\left(\frac{T_{BKT}}{T}-1\right)\Bigg]^{-1}\,\\
\label{C_BoseGas_lowT0}
\earray
where the superscript $^{(0)}$ denotes we are at the lowest order in the
considered approximation.
One can obtain higher order solutions by substituting the expression
in Eq.\,\eqref{solution_theta0} in the logarithmic term of the equation
Eq.\,\eqref{eq_theta(X)} and solve for $\theta(X)$, which will now be
the solution at the first order of approximation, \textit{i.e.} it reads:
\be
\label{solution_theta1}
\theta^{(1)}(X)\,=\,X+\frac{1}{\pi}\ln(2\xi_\mu)+\frac{1}{\pi}\ln\theta^{(0)}(X)\,.
\ee

Following the same procedure sketched above for the zero-th order case,
we obtained the following analytical form for
$\mathcal{C}_0^{\rm Bose}$ at low temperatures at first order approximation:
\barray
\nonumber
&&\mathcal{C}_0^{\rm Bose\,(1)}(T/T_{BKT})\,\simeq\,1+\frac{1}{2}\Bigg\{1-2\ln\Bigg[\frac{2mU}{\hbar^2}\\
\nonumber& & \left(\frac{\hbar^2\xi}{2mU}\right)^{T_{BKT}/T}\Bigg]+W\left(\frac{4\pi mU}{\hbar^2}\left(\frac{\hbar^2\xi}{2mU}\right)^{T_{BKT}/T}\right) \Bigg\}^{-1}\,,\\
\label{C_BoseGas_lowT1}
\earray
where $W(z)$ is the Lambert or product logarithm function.
Higher order solutions may be obtained following the same recipe, but
from the second order case is not possible to write
an analytical expression for $X$ in terms of $T/T_{BKT}$. Therefore, one
can work out only the numerics in order to obtain the low temperature
behaviour of the exponent $\mathcal{C}_0^{\rm Bose\,(j\ge2)}(T/T_{BKT})$. In the present work
the third order approximation has been also investigated, but we envisage no
particular difficulty in going beyond.

\begin{figure}[t]
\includegraphics[width=\columnwidth]{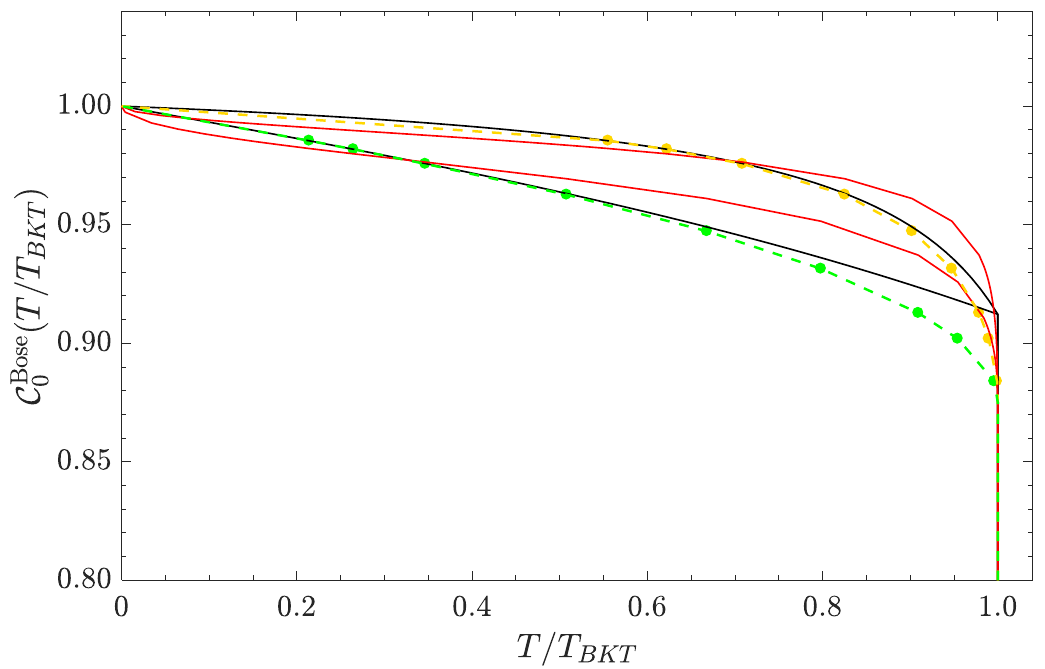}
\caption{Comparison of low temperature predictions for
  $\mathcal{C}_0^{\rm Bose}(T/T_{BKT})$ vs $T/T_{BKT}$ with numerical results for
  the
  two different interactions $\frac{mU}{\hbar^2}= 10^{-12}, 0.25$.
  Bottom green (top yellow) points are the numerical values
  from numerical simulations performed
  in\,\cite{Svistunov2002}, respectively for $\frac{mU}{\hbar^2}= 0.25$
  ($\frac{mU}{\hbar^2}= 10^{-12}$), while dashed lines are drawn
  as guide for the eye.
  Low temperatures predictions from third order approximation
  are reported in black solid lines for two different interaction strengths:
  the line below (above) is for $\frac{mU}{\hbar^2}= 0.25$
  ($\frac{mU}{\hbar^2}= 10^{-12}$) .
  Red solid lines (standing above the black ones for both interaction
  strengths)
  represent the predictions for
  $T\simeq T_{BKT}$ from Eq. (\ref{C_BoseGas_highT}).}
\label{fig5}
\end{figure}

In Fig.\,\ref{fig5} we report the comparison between the low temperature expansions
with the values for $\mathcal{C}_0^{\rm Bose}$ obtained from the
numerical Monte Carlo results of Ref.\,\cite{Svistunov2002}
in the very small interaction limit 
$\frac{mU}{\hbar^2}=10^{-12}$, and for the intermediate interaction case
$\frac{mU}{\hbar^2}=0.25$. The agreement is good up to $3\%$ even for
$T=T_{BKT}$, where
\begin{equation}
  \mathcal{C}_0^{\rm Bose\,(3)}(1)\simeq 0.912\,,
  \label{double}
\end{equation}  
independently of the interaction parameter.
It is important to notice that for smaller values of $U$
the low temperature predictions for the exponent $\mathcal{C}_0^{\rm Bose}$
are valid for a larger range of temperatures, since for very weak
interactions the variable $X$ is very large even at
$T\approx T_{BKT}$. So, decreasing $U$ the range of validity of the low temperature
predictions increase up to a value which becomes increasingly close
to $T_{BKT}$. Indeed, for $\frac{mU}{\hbar^2}=10^{-12}$ the low
$T$ prediction remains reliable up to $T\approx 0.9 T_{BKT}$.

This implies that for $U\rightarrow 0$, \textit{and in practice
  $\frac{mU}{\hbar^2}$ extremely small}, there will be the
above mentioned double jump
phenomenon for the exponent $\mathcal{C}_0^{\rm Bose}$
which will pass near below $T_{BKT}$
from a value close to the quantity in Eq.\,\eqref{double},
\textit{$0.912$}, to $\frac{7}{8}=0.875$ for $T=T_{BKT}$.
Then the second Nelson-Kosterlitz jump
will lead $\mathcal{C}_0^{\rm Bose}$ to pass from $7/8$ to zero. 
It can be seen that there is not appreciable change in this result if
one goes to higher orders of approximation. Despite being not too large in 
absolute value, the first jump should be appreciable in experiments
or simulations, one problem being that one has to 
go possibly to very small values of $\frac{mU}{\hbar^2}$.
We observe that the prediction of the double jump is based on the validity
of the low $T$ expansion and its extension near $T_{BKT}$ for $U$ very small --
and when $T$ is scaled in units of $T_{BKT}$, which in turn depends in $U$.
Therefore it could be that further corrections near $T_{BKT}$ may soften the
first jump, making it a very steep decrease. Notice, that due to
Eq. (\ref{C_2D_general_eta}), the value ${\cal C}_0=0.912$ corresponds
to $\eta=0.176$, which is pretty far from the universal value
$\eta=0.25$, so that going to very small $U$ one should appreciate such
relatively large variation of $\eta$ near $T_{BKT}$.
Further simulations would be extremely useful to better
quantify such steep decrease of $\eta$ close to $T_{BKT}$. 

Interestingly enough, at low temperatures, the Bose gas can be described
by the corresponding results for the XY model. Therefore, posing $\mathcal{C}_0^{XY}=\mathcal{C}_0^{\rm Bose\,(0)}$
\textit{i.e.} equating the low temperature result of the
XY model in Eq.\,\eqref{C_XY_lowT2} to the low temperature result
for the $2D$ Bose gas in Eq.\,\eqref{C_BoseGas_lowT0}
for \textit{any} rescaled temperature $T/T_{BKT}$, one obtains 
the following value for the parameter $\xi$:
\begin{align}
\xi=\frac{1}{2} e^{1+\frac{\pi}{2} ({\cal T}-1)}\,,
\end{align}
where ${\cal T} \equiv 4J/k_B T_{BKT}^{(XY)}$. When the dimensionless interaction strength satisfies the equation
\begin{align}
\frac{mU}{\hbar^2} = \frac{1}{2}e^{1-\frac{\pi}{2}}=0.283\,,
\end{align}
the low $T$ predictions in Eq.\,\eqref{C_BoseGas_lowT0} equals Eq.\,\eqref{C_XY_lowT2}, valid
respectively for the $2D$ Bose gas and the XY model.
Since for the XY model it is
$k_B T_{BKT}^{(XY)}/J=0.893\pm0.001$, one finds
\begin{equation}
  \xi=321 \pm 3,
  \label{321}
\end{equation}  
which should be compared with the Monte Carlo result $\xi=380\pm3$. The
comparison shows that this result (that depends only on the critical temperature
of the $2D$ XY model) is not entirely unreasonable, given  
the non-perturbative nature of the parameter $\xi$ and the well-known
failure of mean-field calculations to determine it
and in general the difficulty of obtaining analytical estimates for it.

Predictions can be made also for $T\simeq T_{BKT}$,
\textit{i.e.} $X\rightarrow 0^+$. We write the function $\theta(X)$ as:
\be
\label{theta_highT}
\theta(X\rightarrow 0)\,=\,bX+\frac{1}{\pi}\ln\left(\frac{\xi}{\xi_\mu}\right)\,,
\ee
where $b$ is a constant to be determined by fitting the values of
$\theta(X)$ for small $X$ coming from Monte Carlo simulations with the law
in Eq.\,\eqref{theta_highT}. It is found $b=1.29 \pm 0.05$. 

For the function $f(X)$ is found
instead\,\cite{Thouless73,Kosterlitz74,Svistunov2002}:
\be
\label{f_highT}
f(X\rightarrow0)\,=\,1+\sqrt{2 \kappa' X}\,,
\ee
with $\kappa'=0.61\pm0.01$. For $\lambda(X)$,
from Eq.\,\eqref{theta_highT}, is simply found that:
\be
\lambda(X\rightarrow0)\,=\,\frac{b-1}{2} \,X\,,
\ee
and therefore, following the same reasoning of the low $T$ case,
from Eq.\,\eqref{T/TBKT_2D_bose} follows that:
\be
\label{X_highT}
X\,=\,\frac{\ln\left(\frac{\hbar^2 \xi }{mU}\right)}{\pi (b-1)}\left(\frac{T_{BKT}}{T}-1\right)\,.
\ee
Finally we can substitute the above expression for $X$ into
Eq.\,\eqref{f_highT} and then into Eq.\,\eqref{C_2D_BoseGas} to obtain an expression
for $\mathcal{C}_0^{\rm Bose}$ for $T\simeq T_{BKT}$ which reads:
\barray
\nonumber
\mathcal{C}_0^{\rm Bose}(T\rightarrow T_{BKT})\,&\simeq&\,1-\frac{1}{8}\Bigg[1+\sqrt{\frac{2\kappa'}{\pi (b-1)}\,\ln\left(\frac{\hbar^2 \xi}{mU}\right)}\\
\label{C_BoseGas_highT}
&&\overline{\left(\frac{T_{BKT}}{T}-1\right)} \Bigg]^{-1}\,.
\earray
We report its behaviour in red solid lines in Fig.\,\ref{fig5}
along with numerical Monte Carlo results of $\mathcal{C}_0^{\rm Bose}$
obtained from Ref.\,\cite{Svistunov2002} for different interactions.
The agreement is good only for values $X\simeq 0$ and the
analytical prediction of Eq. (\ref{C_BoseGas_highT}) gets rapidly worst 
for decreasing temperatures. 

Equating the two behaviours in Eqs.\,\eqref{C_BoseGas_lowT0}
and\,\eqref{C_BoseGas_highT} we can find how the temperature with
which the two curves intersect depends on the dimensionless
interaction parameter $\frac{mU}{\hbar^2}$. Substituting this
expression back to either (\ref{C_BoseGas_lowT0}) or (\ref{C_BoseGas_highT}),
it is found that the value for $\mathcal{C}_0^{\rm Bose}$ at which the two
limiting behaviours intersect is independent on the interaction strength,
and reads:
\barray
\nonumber
\mathcal{C}_0^{\rm Bose\,(0)}\,&=&\,1-\frac{1}{8}\vast[1+\sqrt{\frac{2\kappa'}{\pi(b-1)}}\,\sqrt{5-\ln(2\xi)+\frac{16\kappa'}{\pi (b-1)}}\\
\nonumber
& &\overline{-4\,\sqrt{\frac{2\kappa'}{\pi (b-1)}}\,\sqrt{5-\ln(2\xi)+\frac{8\kappa'}{\pi(b-1)}}}\vast]^{-1}\\
\nonumber 
&&\simeq\,0.914\,.
\earray
This intersection value can also be obtained using the first
order approximation formula $\mathcal{C}_0^{\rm Bose\,(1)}$,
for which one gets $0.915$.

Let now study the scaling exponent ${\cal C}_{k \neq 0}$
for the eigenvalues of the $1$BDM corresponding to non-vaishing momenta.
As in previous Section,
we have to compute the Fourier transform of the symmetrized asymptotic
behaviour of the density matrix, hence:
\barray
\nonumber
\frac{n(k)}{L^2}&\propto& \lim_{L \to \infty} \int_0^{2\pi} e^{i k r \cos(\theta)} d\theta \left[\int_0^{L/2} \frac{dr}{r^{\eta-1}}+\int_{L/2}^L \frac{dr}{(L-r)^{\eta-1}}\right]\\
\nonumber
&&=\,\lim_{L \to \infty}\left[\int_0^{L/2} \frac{J_0(kr)}{r^{\eta-1}}\,dr+\int_{L/2}^L \frac{J_0(kr)}{(L-r)^{\eta-1}}\,dr\right]\,,
\earray
where we passed to polar coordinates symmetrizing on the radial component
as was done for the XY model case, $J_0(x)$ is the Bessel function
of the first kind, and $\eta = \frac{m^2\,k_B T}{2\pi\hbar^2 \rho_s}$
for the weakly interacting Bose gas, while $\eta = \frac{T}{2\pi J_s(T)}$
for the XY model. Focusing only on the first half of the integration
interval\,\cite{footnote} we obtain:
\be
\frac{n(k)}{L^2}\,\propto\,L^{2-\eta}\,_1F_2\left(1-\frac{\eta}{2}; 1, 2-\frac{\eta}{2}; -\frac{\pi^2 l^2}{4}\right)\,,
\ee
where we used $k\,L=2\,\pi\,l$ with $l\in\mathbb{N}$.
Expanding the hypergeometric function for large $l$ and
focusing only the leading term, we obtain finally:
\be
\frac{n(k)}{L^2}\,\propto\,L^{2-\eta}\,l^{\eta-2}\,\propto\,N^0\,,
\ee
where in the last proportion we wrote $l\propto L$ in order that
$k$ remains finite in the thermodynamic limit and $L\propto\sqrt{N}$,
since the density $n=N/L^2$ is fixed. Therefore we simply read
\be
\label{C_2D_kneq0}
\mathcal{C}_{k\neq0}(T)\,=\,0\,,
\ee
both for the XY and two--dimensional Bose gas systems for zero and
finite temperatures.

\section{Conclusions}
\label{concl}

The goal of the present paper has been to characterize
the off-diagonal long-range order (ODLRO)
properties of interacting bosons at finite
temperatures through the study of the eigenvalues' scaling of the one--body density matrix ($1$BDM) vs the number of particles $N$.
For translational invariant systems, denoting by $\lambda_k$ the eigenvalues
of the ($1$BDM) and by $\lambda_0$ the largest among them,
one can define the scaling exponents ${\cal C}_k$ from the relation
$\lambda_k \sim N^{{\cal C}_k}$. The exponents ${\cal C}_k$ depend on the
temperature $T$ and on the strength of the interaction
(which we assume short-ranged), and as well on the dimension $D$.
According the Penrose-Onsager criterion, $\mathcal{C}_0=1$ corresponds to
ODLRO, while at variance the opposite limit $\mathcal{C}_0=0$
corresponds to the single-particle occupation of the natural orbital
associated to $\lambda_0$. The intermediate case,
$0<\mathcal{C}_0<1$, is associated for translational invariant systems
to the power-law decaying of non-connected correlation functions and
it can be seen as identifying quasi-long-range order.

After introducing some basic definitions and properties of
the $1$BDM, we discussed how to obtain the exponents $\mathcal{C}_k$
directly from the large distance behaviour of the $1$BDM.
The ODLRO in the three--dimensional case for temperatures below the
Bose-Einstein critical temperature has been described,
as well as quasi-long-range order in the one-- and two--dimensional Bose gases
for different interactions and temperatures, discussing the connection
of the Mermin-Wagner theorem with the occurrence
of mesoscopic condensation.
We showed that in $1D$ it is $\mathcal{C}_0=0$
for non-vanishing temperature, while in $3D$ $\mathcal{C}_0=1$
($\mathcal{C}_0=0$) for temperatures smaller (larger) than
the Bose-Einstein critical temperature. We then focused on the
two--dimensional case.
We presented the application of our
methods to the XY and Villain models,
where ODLRO is translated as a fully magnetization of the system, and to the $2D$
Bose gases. A universal jump of the power $\mathcal{C}_0$
from $\frac{7}{8}$ to $0$ is
found at the Berezinskii--Kosterlitz--Thouless temperature $T_{BKT}$,
reflecting the universal jump for the superfluid stiffness.
The dependence of $\mathcal{C}_0$ between $T=0$
(at which $\mathcal{C}_0=1$)
and $T_{BKT}$ is studied in the different models.
We found a weak dependence
of it when the reduced temperature $T/T_{BKT}$ is used.
An estimate for the (non-perturbative) parameter $\xi$ entering the
equation of state of the $2D$ Bose gases was obtained using low temperature
expansions and compared with the Monte Carlo result. We also unveiled
a ``double jump''-like behaviour for $\mathcal{C}_0$, and
correspondingly of the anomalous dimension $\eta$,
right below $T_{BKT}$ in the limit of vanishing interactions.
When the dimensionless parameter $mU/\hbar^2$ is very small,
the validity region of the low-temperature expansions enlarges towards
$T_{BKT}$ as soon as that $mU/\hbar^2$ decreases. When such regime is reached,
then $\mathcal{C}_0$ tends to the value $\approx 0.912$,
and again moving towards
$T_{BKT}$ from below it abruptly
(or, at least, in a very steep way) decreases to the universal value $7/8$,
then jumping again to $0$. We presented a
detailed discussion of the weakly interacting regime and we commented 
how the double jump behaviour could be appreciable for very low
values of the parameter $mU/\hbar^2$. Then we
analyzed the behaviour of
$\mathcal{C}_{k\neq0}$, finding that in none of the cases presented
there is quasi-fragmentation, 
\textit{i.e.} $\mathcal{C}_{k\neq0}=0$.

Our investigation is based both on the homogeneity of space and the
thermodynamic limit, therefore will be interesting to study in a future
work whether adding a confining external potential could change our
predictions and how finite number of particles affects the results. Moreover,
it would be of interest to consider long-range interactions\,\cite{Defenu19}
and the presence of
disorder, where rigorous results are available in
literature\,\cite{Seiringer12,Koneberg15}. We also mention that for
$2D$ anyonic gases, despite the presence of a considerable literature, see
\textit{e.g.}\,\cite{Khare05,Mancarella13,Lundholm13,Ouvry18} and refs.
therein, to the best of our knowledge no results for the scaling exponents
${\cal C}_k(T)$ are available at date. 

{\it Acknowledgements:} We thank T. Enss, L. Lepori, D. Lundholm and
I. Nandori for discussions and J. Yngvason and M. Hasenbusch
for useful correspondence. A.T.
acknowledge kind hospitality at ``Mathematical physics of anyons and
topological states of matter'', taking place in
Nordita, Stockholm (Sweden), March 2019,
where parts of present work have been fruitfully
discussed with participants to the conference.  This work is supported by the Deutsche Forschungsgemeinschaft (DFG, German Research Foundation) under Germany’s Excellence Strategy ``EXC-2181/1-390900948'' (the Heidelberg STRUCTURES Excellence Cluster). N.D. and A.T acknowledge support
from the CNR/MTA Italy-Hungary 2019-2021 Joint Project ''Strongly interacting systems in confined geometries''.


\begin{thebibliography}{XX}
\bibitem{footnote}
   We observe that symmetrizing the density matrix adding the
  mirrored term in the region $L/2$ to $L$ serves to have
  a positive and real result for the momentum distribution
  and occupation numbers, but it does not
  affect the scaling of the $\lambda_k$ eigenvalues in terms of $L$.
\end{thebibliography}
\end{document}